\documentclass[aps,showpacs,preprintnumbers,amsmath,amssymb]{revtex4}

\usepackage{float}
\usepackage{graphics,epsfig}
\usepackage{graphicx}
\usepackage{dcolumn}
\usepackage{bm}

\begin{document}
\baselineskip=0.8 cm
\title{{\bf Greybody Factors for Rotating Black Holes on Codimension-2 Branes}}
\author{Songbai Chen}
\email{csb3752@163.com} \affiliation{Department of Physics, Fudan
University, 200433 Shanghai
 \\ Institute of Physics and  Department of Physics,
Hunan Normal University,  Changsha, 410081 Hunan  }

\author{Bin Wang}
\email{wangb@fudan.edu.cn} \affiliation{Department of Physics,
Fudan University, 200433 Shanghai}

\author{ Rukeng Su}
\email{rksu@fudan.ac.cn}
 \affiliation{China Center of Advanced Science and Technology (World Laboratory),
P.B.Box 8730, 100080 Beijing
\\ Department of Physics, Fudan University, 200433 Shanghai}

\author{W.-Y Pauchy Hwang}
\affiliation{Department of Physics, National Taiwan University,
106 Taipei}

 \vspace*{0.2cm}
\begin{abstract}
\baselineskip=0.6 cm
\begin{center}
{\bf Abstract}
\end{center}

We study the absorption probability and Hawking radiation of the
scalar field in the rotating black holes on codimension-2 branes.
We find that finite brane tension modifies the standard results in
Hawking radiation if compared  with the case when brane tension is
completely negligible.  We observe that the rotation of the black
hole brings richer physics. Nonzero angular momentum triggers the
super-radiance which becomes stronger when the angular momentum
increases. We also find that rotations along different angles
influence the result in absorption probability and Hawking
radiation.  Compared with the black hole rotating orthogonal to
the brane, in the background that black hole spins on the brane,
its angular momentum brings less super-radiance effect and the
brane tension increases the range of frequency to accommodate
super-radiance. These information can help us know more about the
rotating codimension-2 black holes.

\end{abstract}

 \maketitle
\newpage
\section{Introduction}

In braneworld scenarios it is known that large extra dimensions
can lower the fundamental scale of gravity down to the order of
TeV. This brings the possibility of creating microscopic black
holes in high energy experiments such as the forthcoming Large
Hadron Collider (LHC) \cite{xx} (for a recent review on this topic
and a complete list of references please refer to \cite{xx1}).
After their creation, these microscopic black holes will decay
quickly through the emission of Hawking radiation. Hawking
radiation serves as the chief arena for studying the quantum
gravity and disclosing the signature of extra dimensions
\cite{1}-\cite{11}. Other attempts on detecting the extra
dimension have been investigated in the perturbations around
braneworld black holes \cite{12}-\cite{16}.

Most examinations for the extra dimension through Hawking
radiations and the wave dynamics are concentrated on braneworld
black holes with zero brane tension. However the nonzero tension
on the brane is not trivial, since it can curve the brane as well
as the bulk. In general it is very hard to obtain exact solutions
of higher-dimensional Einstein equations describing black holes on
the brane with tension. Recently, a metric describing a black hole
located on a three-brane with finite tension, embedded in locally
flat six-dimensional spacetime was constructed in \cite{17}.
Hawking radiation \cite{7} and perturbations around this black
hole \cite{14,16} have been studied subsequently, where
modifications due to the finite brane tension and  imprints of
extra dimensions have been examined.

Further progress has been made by generalizing the Schwarzschild
like solution of \cite{17} to rotating black holes on a
codimension-2 brane \cite{18}. Including the rotation will bring
richer physics.  In Hawking radiation the amplifications in various
properties due to the rotation have been shown in \cite{9,10,11}
where the brane tension was zero. It is of great interest to examine
the evaporation of the rotating black holes on codimension-2 brane
and investigate the physics brought by the rotation and the brane
tension.

The general metric of a small rotating black holes on a
codimension-2 brane reads \cite{18}
\begin{eqnarray}
ds^2&=&-(1-\frac{\mu r}{\Pi F})dt^2+\frac{2\mu r}{\Pi
F}(a_1\mu^2_1 d\phi_1+ba_2\mu^2_2 d\phi_2)dt+\frac{F}{1-\frac{\mu
r}{\Pi}}dr^2+r^2(d\alpha^2+d\mu^2_1+d\mu^2_2+\mu^2_1d\phi^2_1+b^2\mu^2_2d\phi^2_2)\nonumber\\
&+&a^2_1[d\mu^2_1+\mu^2_1(1+\frac{\mu r}{\Pi
F}\mu^2_1)d\phi^2_1]+a^2_2[d\mu^2_2+b^2\mu^2_2(1+\frac{\mu r}{\Pi
F}\mu^2_2)d\phi^2_2]+2ba_1a_2\frac{\mu r}{\Pi
F}\mu^2_1\mu^2_2d\phi_1d\phi_2, \label{metric0}
\end{eqnarray}
with
\begin{eqnarray}
\Pi=\sum^2_{i=1}(r^2+a^2_i),\;\;\;\;\;\;\;\;\;\;\;
F=\sum^2_{i=1}\frac{r^2\mu^2_i}{r^2+a^2_i}+\alpha^2.
\end{eqnarray}
Here $a_1$, $a_2$ are two angular momentum parameters. The two
direction cosines, $\mu_i$ and the coordinate $\alpha$ are
constrained by
\begin{eqnarray}
\mu^2_1+\mu^2_2+\alpha^2=1.
\end{eqnarray}
The quantity $\mu=r^3_s/b$ is proportional to the mass of the
black hole, where $r_s$ is the horizon radius of the usual
six-dimensional Schwarzschild black hole. The parameter $b$ is
related to the brane tension $\lambda$ by $b
=1-\frac{\lambda}{2\pi M^4_*}$ with $0<b\leq 1$, where $ M_*$ is
the fundamental mass scale of six-dimensional gravity.

There are two different choices of angles through which the hole
can rotate. Choosing the rotation axis to be $\phi_2$, by setting
$a_1=0, a_2=a$, and after doing rescale $\phi_2\rightarrow
b\phi_2$ and substituting  $\phi_1=\phi$, $\phi_2= \psi$, and
\begin{eqnarray}
\mu_1=\sin{\theta},\;\;\;\;\;\;\;\;\;\;\;
\mu_2=\cos{\theta}\sin{\chi},
\;\;\;\;\;\;\;\;\;\alpha=\cos{\theta}\cos{\chi},
\end{eqnarray}
into equation (\ref{metric0}),  we can obtain the metric for the
black hole spinning orthogonal to the brane with its angular
momentum pointing along the brane
\begin{eqnarray}
ds^2&=&-(1-\frac{\mu }{r \rho^2})dt^2+\frac{2\mu a
}{r\rho^2}\sin{\theta}^2dtd\phi
+\frac{\rho^2}{\Delta}dr^2+\rho^2d\theta^2\nonumber\\
&+&\sin{\theta}^2(r^2+a^2+\frac{\mu
a^2\sin{\theta}^2}{r\rho^2})d\phi^2+r^2\cos{\theta}^2(d\chi^2+b^2\sin{\chi}^2d\psi^2),
\label{metric1}
\end{eqnarray}
where $\rho^2=r^2+a^2\cos{\theta}^2$ and $\Delta=r^2+a^2-\mu/r$.
The Hawking temperature and the angular velocity of the horizon
can be expressed as
\begin{eqnarray}
T_H=\frac{3r^2_H+a^2}{4\pi r_H(r^2_H+a^2)},
\end{eqnarray}
and
\begin{eqnarray}
\Omega_H=\frac{a}{(r^2_H+a^2)}\label{ji1}.
\end{eqnarray}
Due to the presence of $b$ the radius of horizon $r_H$ is larger
than $r_s$ \cite{18}, the angular velocity at the horizon is smaller
than that in the tensionless case.

Choosing $\phi_1$ as the rotation axis, the hole would be spinning
on the brane with its angular momentum orthogonal to the brane,
whose metric has the form
\begin{eqnarray}
ds^2&=&-(1-\frac{\mu }{r \rho^2})dt^2+\frac{2\mu a
b}{r\rho^2}\sin{\theta}^2dtd\phi
+\frac{\rho^2}{\Delta}dr^2+\rho^2d\theta^2\nonumber\\
&+&b^2\sin{\theta}^2(r^2+a^2+\frac{\mu
a^2\sin{\theta}^2}{r\rho^2})d\phi^2+r^2\cos{\theta}^2(d\chi^2+\sin{\chi}^2d\psi^2).
\label{metric2}
\end{eqnarray}
The Hawking temperature is the same as that in the previous case,
but the angular velocity at the horizon becomes
\begin{eqnarray}
\Omega_H=\frac{a}{b (r^2_H+a^2)}\label{ji2}.
\end{eqnarray}
It is clear that this angular velocity is larger than that in the
tensionless case because the radius of the black hole horizon
$r_H\sim b^{-1/3}r_s$, and thus the denominator in  equation (9)
$b(r^2_H+a^2)\sim b^{1/3}r^2_s+ba^2<r^2_s+a^2$.

We will discuss the evaporation of scalar field in six-dimensional
black holes pierced by a tense 3-brane rotating orthogonal to the
brane and along the brane. We will calculate the corresponding
absorption probabilities and luminosity of Hawking radiation
analytically by employing the matching technique of combining the
far-field and near-horizon parts of solutions in the low energy
and angular momentum limit.

The organization of the paper is as follows: in the following
section we will derive the master equation in rotating black holes
on codimension-2 branes. In Sec.III we will present the solution in
the low energy and low angular momentum limit by using the matching
technique. In Sec.IV, we derive the absorption probability and the
luminosity of Hawking radiation. Finally in the last section we
present our conclusions.

\section{The Master Equation in Rotating Black Holes on Codimension-2 Branes}

The equation of motion for a massless scalar particle propagating
in the curved spacetime is described by
\begin{eqnarray}
\frac{1}{\sqrt{-g}}\partial_{\mu}(\sqrt{-g}g^{\mu\nu}\partial_{\nu})
\Phi(t,r,\theta,\phi,\chi,\psi)=0,\label{WE}
\end{eqnarray}
where $\Phi(t,r,\theta,\phi,\chi,\psi)$ is the scalar field.
Separating the scalar field into
$\Phi(t,r,\theta,\phi,\chi,\psi)=e^{-i\omega
t+im\phi+i\eta\psi}R(r)S(\theta)\Gamma(\chi)$, we can obtain the
radial and angular equations for the metric (\ref{metric1})
\begin{eqnarray}
\frac{1}{r^2}\frac{d}{dr}\bigg[r^2\Delta\frac{d R(r)}{dr}\bigg]
+\bigg[\frac{K^2_1}{\Delta}+2am\omega-a^2\omega^2-E_{lmj1}-\frac{\Lambda(j,\eta)a^2}{r^2}\bigg]R(r)=0,\label{radial}
\end{eqnarray}
\begin{eqnarray}
\frac{1}{\sin{\theta}\cos{\theta}^2}\frac{d}{d\theta}\bigg[\sin{\theta}\cos{\theta}^2
\frac{d}{d\theta}\bigg]S(\theta)
+\bigg[\omega^2a^2\cos{\theta}^2-\frac{m^2}{\sin^2{\theta}}
-\frac{\Lambda(j,\eta)}{\cos^2{\theta}}-E_{lmj1}\bigg]S(\theta)=0,\label{angd1}
\end{eqnarray}
\begin{eqnarray}
\frac{1}{\sin{\chi}}\frac{d}{d\chi}\bigg[\sin{\chi}\frac{d}{d\chi}\Gamma(\chi)\bigg]
+\bigg[\Lambda(j,\eta)-\frac{\eta^2}{b^2\sin^2{\chi}}\bigg]\Gamma(\chi)=0,\label{angd3}
\end{eqnarray}
with
\begin{eqnarray}
K_1=\omega (r^2+a^2)-am,
\end{eqnarray}
and $\eta$ and $\Lambda(j,\eta)$ are angular eigenvalue.

Similarly, in the background metric (\ref{metric2}) radial and
angular equations can be expressed as
\begin{eqnarray}
\frac{1}{r^2}\frac{d}{dr}\bigg[r^2\Delta\frac{d R(r)}{dr}\bigg]
+\bigg[\frac{K^2_2}{\Delta}+\frac{2am\omega}{b}-a^2\omega^2-E_{lmj2}-\frac{j(j+1)a^2}{r^2}\bigg]R(r)=0,\label{radia2}
\end{eqnarray}
\begin{eqnarray}
\frac{1}{\sin{\theta}\cos{\theta}^2}\frac{d}{d\theta}\bigg[\sin{\theta}\cos{\theta}^2
\frac{d}{d\theta}\bigg]S(\theta)
+\bigg[\omega^2a^2\cos{\theta}^2-\frac{m^2}{b^2\sin^2{\theta}}-\frac{j(j+1)}
{\cos^2{\theta}}-E_{lmj2}\bigg]S(\theta)=0,\label{angd2}
\end{eqnarray}
with
\begin{eqnarray}
K_2=\omega (r^2+a^2)-\frac{am}{b}.
\end{eqnarray}

We limit ourselves to the case where $\omega a\ll1 $ and the
deviation of the parameter $b$ from unity is very small which is
physically justified for small brane tension. And then we can
adopt the perturbation theory to calculate eigenvalues of angular
equations (\ref{angd1}), (\ref{angd3}) and (\ref{angd2}). This
perturbation method was first used in \cite{14} and was supported
in \cite{16}. As did in \cite{14}, we find the angular eigenvalue
of Eq. (\ref{angd3})
\begin{eqnarray}
\Lambda(j,\eta)=j(j+1)+\frac{\eta(2j+1)(1-b^2)}{2b^2}.
\end{eqnarray}
The zeroth-order eigenfunctions of (\ref{angd1}) and (\ref{angd2})
can be given in terms of the Jacobi polynomials
\begin{eqnarray}
S_0(\theta)=(\sin{\theta})^{|m|}\cos{\theta}^{|j|}
P\bigg(\frac{l-j-m}{2},j+\frac{1}{2},m; 1-2\cos^2{\theta}\bigg),
\end{eqnarray}
and then angular eigenvalues of Eqs. (\ref{angd1}) and
(\ref{angd2}) can be expressed respectively
\begin{eqnarray}
E_{lmj1}&=&l(l+3)+\frac{\eta(2j+1)(1-b^2)}{2b^2}\frac{[2(j+m)+1](2j+3)+2(l-j-m)(l+j+m+3)}{(2j+3)(2j-1)}
\nonumber\\
&+&a^2\omega^2\frac{2j(j-1)+2l(l+1)-2m^2+4(l+j)+3}{(2l+5)(2l+1)},
\end{eqnarray}
\begin{eqnarray}
E_{lmj2}&=&l(l+3)+\frac{m[2(j+m)+1](1-b^2)}{2b^2}+a^2\omega^2\frac{2j(j-1)+2l(l+1)-2m^2+4(l+j)+3}{(2l+5)(2l+1)},
\end{eqnarray}
where $l$, $m$, $j$ and $\eta$ are restricted by
\begin{eqnarray}
l\geq (j+|m|),\;\;\;\;\;\;\;\; j\geq |\eta|\;\;\;\;\;\;\;\;
\text{and} \;\;\;\;\;\;\;\; \frac{l-(j+m)}{2}\in
\{0,\mathbb{Z}^+\}.
\end{eqnarray}
In the limit $b\rightarrow 1$, the eigenvalue $E_{lmj1}$ is
identical to $E_{lmj2}$ and returns to that in the six dimensional
rotating black hole spacetime without brane tension.

\vspace*{0.2cm}
\section{Greybody Factor in the Low-Energy Regime}
Now we provide an analytic solution of the radial equation using
the matching technique. We first derive the solution in the near
horizon regime, then derive the solution in the far field limit.
Finally we stretch and match the two solutions in an intermediate
region. In this way we can construct the analytic expression in
the low energy and low angular momentum approximation for the
radial part of the field valid throughout the entire spacetime.
This analytic approximation has been employed in \cite{9,10,11}.

Let us first focus on the near-horizon regime. In order to
translate radial equations (\ref{radial}) and (\ref{radia2}) into
the form of the known differential equation, we make the following
change of the variable
\begin{eqnarray}
r\rightarrow f=\frac{\Delta}{r^2+a^2}\Rightarrow \frac{d
f}{dr}=(1-f)r\frac{A}{r^2+a^2},
\end{eqnarray}
where $A=3+a^2/r^2$. Then radial equations (\ref{radial}) and
(\ref{radia2}) can be rewritten in a unified form
\begin{eqnarray}
f(1-f)\frac{d^2R(f)}{d f^2}+(1-D_*f)\frac{d R(f)}{d f}
+\bigg[\frac{K^2_{*,i}}{A(r_H)^2(1-f)f} -\frac{(E_{lmj
i}-P_i)(r^2_H+a^2)}{r^2_HA(r_H)^2(1-f)}\bigg]R(f)=0,\label{r1}
\end{eqnarray}
with $i=1,2$ corresponding to new radial equations in the
background of (\ref{metric1}) and (\ref{metric2}) respectively.
\begin{eqnarray}
D_*=1-\frac{4a^2r^2_H}{(3r^2_H+a^2)^2},
\end{eqnarray}
and
\begin{eqnarray}
K_{*,1}=\omega
(r_H+\frac{a^2}{r_H})-\frac{am}{r_H},\;\;\;\;\;\;\;\;
P_1=2am\omega-a^2\omega^2-\frac{\Lambda(j,\eta)a^2}{r^2_H},
\end{eqnarray}
\begin{eqnarray}
K_{*,2}=\omega
(r_H+\frac{a^2}{r_H})-\frac{am}{br_H},\;\;\;\;\;\;\;\;
P_2=\frac{2am\omega}{b}-a^2\omega^2-\frac{j(j+1)a^2}{r^2_H}.
\end{eqnarray}
for black holes on codimension-2 branes rotating orthogonal to the
brane or on the brane, respectively.  Making the field
redefinition $R(f)=f^{\alpha}(1-f)^{\beta}F(f)$, we can rewrite
equation (\ref{r1}) in the form of a hypergeometric equation
\begin{eqnarray}
f(1-f)\frac{d^2F(f)}{d
f^2}+[c-(1+\tilde{a}_i+\tilde{b}_i)f]\frac{d F(f)}{d
f}-\tilde{a}_i\tilde{b}_i F(f)=0,\label{r2}
\end{eqnarray}
where
\begin{eqnarray}
\tilde{a}_i=\alpha_i+\beta_i+D_*-1,\;\;\;\;\;\;\;\;\;\;
\tilde{b}_i=\alpha_i+\beta_i,\;\;\;\;\;\;\;\;\;\;\;\;\;
c_i=1+2\alpha_i.
\end{eqnarray}
Due to the constraint from the coefficient of $F(f)$, the power
coefficients $\alpha_i$ and $\beta_i$ must satisfy the
second-order algebraic equations
\begin{eqnarray}
\alpha^2_i+\frac{K^2_{*,i}}{A(r_H)^2}=0,
\end{eqnarray}
and
\begin{eqnarray}
\beta^2_i+\beta_i(D_*-2)+\frac{K^2_{*,i}}{A(r_H)^2}-
\frac{(E_{lmji}-P_i)(r^2_H+a^2)}{r^2_HA(r_H)^2}=0.
\end{eqnarray}
Solving these two equations, we obtain solutions for parameters
$\alpha$ and $\beta$
\begin{eqnarray}
&&\alpha_{i\pm}=\pm \frac{iK_{*,i}}{A(r_H)},\\
&&\beta_{i\pm}=\frac{1}{2}\bigg[(2-D_*)\pm\sqrt{(D_*-2)^2-\frac{4K^2_{*,i}}{A(r_H)^2}
+\frac{4(E_{lmji}-P_i)(r^2_H+a^2)}{r^2_HA(r_H)^2}} \;\bigg].
\end{eqnarray}
Thus, the general solution of master equations (\ref{radial}) and
(\ref{radia2}) near the horizon can be expressed as
\begin{eqnarray}
R_{iNH}(f)=A_{i-}f^{\alpha_i}(1-f)^{\beta_i}F(\tilde{a_i},\tilde{b_i},c_i;
f)+A_{i+}f^{-\alpha}(1-f)^{\beta}F(\tilde{a_i}-c_i+1,\tilde{b_i}-c_i+1,2-c_i;
f),\label{s0}
\end{eqnarray}
where $A_{i\pm}$ are arbitrary constants. Near the horizon,
$r\rightarrow r_H$, and $f\rightarrow 0$, the solution (\ref{s0})
can be reduced to
\begin{eqnarray}
R_{iNH}(f)=A_{i-}f^{\pm iK_{*,i}/A(r_H)}+A_{i+}f^{\mp
iK_{*,i}/A(r_H)}=A_{i-}e^{\pm ik_iy}+A_{i+}e^{\mp
ik_iy},\label{s1}
\end{eqnarray}
with
\begin{eqnarray}
k_1=\omega-\frac{am}{r^2_H+a^2},\;\;\;\;\;\;\;k_2=\omega-\frac{am}{b(r^2_H+a^2)},
\end{eqnarray}
for different rotating angles. Here $y$ is the tortoise-like
coordinate, which can be expressed as
\begin{eqnarray}
y=\frac{(r^2_H+a^2)\ln{f}}{r_HA(r_H)}.
\end{eqnarray}
According to the boundary condition that no outgoing mode exists
near the horizon, we choose $\alpha_i=\alpha_{i-}$ and $A_{i+}=0$.
Thus the asymptotic near horizon solution has the form
\begin{eqnarray}
R_{iNH}(f)=A_{i-}f^{\alpha_i}(1-f)^{\beta_i}F(\tilde{a_i},
\tilde{b_i}, c_i; f).
\end{eqnarray}
Moreover, the boundary condition also demands that near the
horizon the hypergeometric function $F(\tilde{a_i}, \tilde{b_i},
c_i; f)$ must be convergent, i.e.
$Re(c_i-\tilde{a_i}-\tilde{b_i})> 0$, which implies that we must
choose $\beta_i=\beta_{i-}$.

Now, let us turn to the far field region, where equations
(\ref{radial}) and (\ref{radia2}) take the form
\begin{eqnarray}
\frac{d^2R_{FF}(r)}{d r^2}+\frac{4}{r}\frac{dR_{FF}(r)}{d
r}+\bigg[\omega^2-\frac{E_{lmji}+a^2\omega^2}{r^2}\bigg]R_{FF}(r)=0.
\end{eqnarray}
Obviously, this is a Bessel equation. General solutions of radial
master equations (\ref{radial}) and (\ref{radia2}) in the far
field region can be expressed as
\begin{eqnarray}
R_{iFF}(r)=\frac{1}{\sqrt{r}}\bigg[B_{i1}J_{\nu_i}(\omega
r)+B_{i2}Y_{\nu_i} (\omega r)\bigg],\label{rf}
\end{eqnarray}
where $J_{\nu_i}(\omega r)$ and $Y_{\nu_i}(\omega r)$ are the
first and second kind Bessel functions, $\nu_
i=\sqrt{E_{lmji}+a^2\omega^2+9/4}$. $B_{i1}$ and $B_{i2}$ are
integration constants.

In order to match the near horizon and far field solutions in the
intermediate zone, we must stretch the near horizon solution to
the large value of the radial coordinate. As in Refs.
\cite{9,10,11}, at first we change the argument of the
hypergeometric function of the near-horizon solution from $f$ to
$1-f$ by using the relation
\begin{eqnarray}
R_{iNH}(f)&=&A_-f^{\alpha_i}(1-f)^{\beta_i}
\bigg[\frac{\Gamma(c-i)\Gamma(c_-\tilde{a_i}-\tilde{b_i})}{\Gamma(c_i-\tilde{a_i})\Gamma(c_i-\tilde{b_i})}
F(\tilde{a_i}, \tilde{b_i}, \tilde{a_i}+\tilde{b_i}-c_i+1; 1-f)\nonumber\\
&+&(1-f)^{c_i-\tilde{a_i}-\tilde{b_i}}\frac{\Gamma(c_i)\Gamma(\tilde{a_i}+\tilde{b_i}-c_i)}{\Gamma(\tilde{a_i})
\Gamma(\tilde{b_i})} F(c_i-\tilde{a_i}, c_i-\tilde{b_i},
c_i-\tilde{a_i}-\tilde{b_i}+1; 1-f)\bigg].\label{r2}
\end{eqnarray}
In the limit $r\gg r_H$, the function $(1-f)$ can be approximated
by
\begin{eqnarray}
1-f=\frac{\mu}{r}\frac{1}{r^2+a^2}\sim\frac{r_H(r^2_H+a^2)}{r^3},
\end{eqnarray}
and the near-horizon solution (\ref{r2}) can be simplified further
as
\begin{eqnarray}
R_{iNH}(r)\simeq
A_{i1}r^{-3\beta_i}+A_{i2}r^{3(\beta_i+D_*-2)}\label{rn2},
\end{eqnarray}
with
\begin{eqnarray}
A_{i1}=A_{i-}\frac{\Gamma(c_i)\Gamma(c_i-\tilde{a_i}-\tilde{b_i})}
{\Gamma(c_i-\tilde{a_i})\Gamma(c_i-\tilde{b_i})}[r_H(r^2_H+a^2)]^{\beta_i},\label{rn3}
\end{eqnarray}
\begin{eqnarray}
A_{i2}=A_{i-}\frac{\Gamma(c_i)\Gamma(\tilde{a_i}+\tilde{b_i}-c_i)}
{\Gamma(\tilde{a_i})\Gamma(\tilde{b_i})}[r_H(r^2_H+a^2)]^{-(\beta+D_*-2)}.\label{rn4}
\end{eqnarray}
In the limit $r\rightarrow 0$, $R_{iFF}(r)$ in equation (\ref{rf})
becomes
\begin{eqnarray}
R_{iFF}(r)\simeq\frac{B_{i1}(\frac{\omega
r}{2})^{\nu_i}}{\sqrt{r}\;\Gamma(\nu_i+1)}
-\frac{B_{i2}\Gamma(\nu_i)}{\pi \sqrt{r}\;(\frac{\omega
r}{2})^{\nu_i}}.
\end{eqnarray}
And then comparing it with equation (\ref{rn2}), we will obtain
two relations between $A_{i1}$ and $B_{i1},\;B_{i2}$ in the limit
$\omega r_H\ll1$.  Then making use of equations (\ref{rn3}) and
(\ref{rn4}) and removing $A_{i-}$, we find that the constraint for
$B_{i1},\; B_{i2}$ is given by
\begin{eqnarray}
B_i\equiv\frac{B_{i1}}{B_{i2}}&=&-\frac{1}{\pi}\bigg[\frac{2
}{\omega r^{1/3}_H(r^2_H+a^2)^{1/3}}\bigg]^{2l+3}\sqrt{E_{lmji}+a^2\omega^2+9/4}\nonumber\\
&\times& \frac{\Gamma^2(\sqrt{E_{lmj}+a^2\omega^2+9/4})
\Gamma(c_i-\tilde{a_i}-\tilde{b_i})\Gamma(\tilde{a_i})\Gamma(\tilde{b_i})}
{\Gamma(\tilde{a_i}+\tilde{b_i}-c_i)\Gamma(c_i-\tilde{a_i})\Gamma(c_i-\tilde{b_i})}.
\label{BB}
\end{eqnarray}
In the asymptotic region $r\rightarrow \infty$, the solution in
the far field can be expressed as
\begin{eqnarray}
R_{iFF}(r)\simeq
\frac{B_{i1}+iB_{i2}}{2\sqrt{2\pi\omega}\;r}e^{-i\omega r}+
\frac{B_{i1}-iB_{i2}}{2\sqrt{2\pi \omega}\;r}e^{i\omega r}=
A^{(\infty)}_{i\;in}\frac{e^{-i\omega
r}}{r}+A^{(\infty)}_{i\;out}\frac{e^{i\omega r}}{r}.\label{rf6}
\end{eqnarray}
The absorption probability can be calculated by
\begin{eqnarray}
|\mathcal{A}_{lmji}|^2=1-\bigg|\frac{A^{(\infty)}_{i\;out}}{A^{(\infty)}_{i\;in}}\bigg|^2
=1-\bigg|\frac{B_i-i}{B_i+i}\bigg|^2=\frac{2i(B^*_i-B_i)}{B_iB^*_i+i
(B^*_i-B_i)+1}.\label{GFA}
\end{eqnarray}

Combining the above result and the expression $B_i$ given in
equation (\ref{BB}), we can examine properties of absorption
probability for the massless scalar field in two kinds of
classical rotating black holes on codimension-2 branes in the
low-energy and low-angular momentum limit.

\vspace*{0.2cm}
\section{ The absorption probability and Hawking
radiation in the rotating black holes on codimension-2 branes}

With solutions obtained above, we are now in a position to compute
the absorption probability and discuss Hawking radiation of black
holes on codimension-2 branes spinning orthogonal to the brane and
spinning on the brane, respectively. Recently numerical study on
the emission of scalar fields into the bulk from a six-dimensional
tensional black hole rotating orthogonal to the brane was also
proposed in \cite{19}.

In Fig.1, we examine the influence of the brane tension on the
absorption probability. We plot the absorption probability for the
first partial waves ($l=0, m=0, j=0$) by fixing $a=0.4$. It is
clearly shown that the absorption probability decreases with the
increase of $b$ (decrease of the brane tension). In Fig.2 we fix
 $b$ (with constant brane tension) and exhibit the dependence of the
absorption probability on the angular momentum. It shows that with
the increase of the angular momentum, the absorption probability
decreases. The main reason is that in the low-energy limit the
absorption probability $|\mathcal{A}(l=0,m=0,j=0)|^2\sim \omega^4
r^4_H$. When parameters $a$ and $b$ increase, the radius of the
black hole event horizon decreases. The dependence of the
absorption on the brane tension and angular momentum of the black
hole does not differ much for the six-dimensional tensional black
holes rotating along different angles.

Fig.2 shows the dependence of the absorption probability on the
angular index. We see the suppression of $|\mathcal{A}|^2$ as the
values of the angular index increase. This phenomenon explains
that the first partial wave dominates over all others, which has
also been observed in black hole cases when there is no brane
tension \cite{9,10}. Moreover, we also observe that the absorption
probability in the black hole (\ref{metric1}) depends on the
angular index $\eta$. With the increase of the index $\eta$, the
absorption probability decreases.
\begin{figure}[ht]
\begin{center}
\includegraphics[width=8cm]{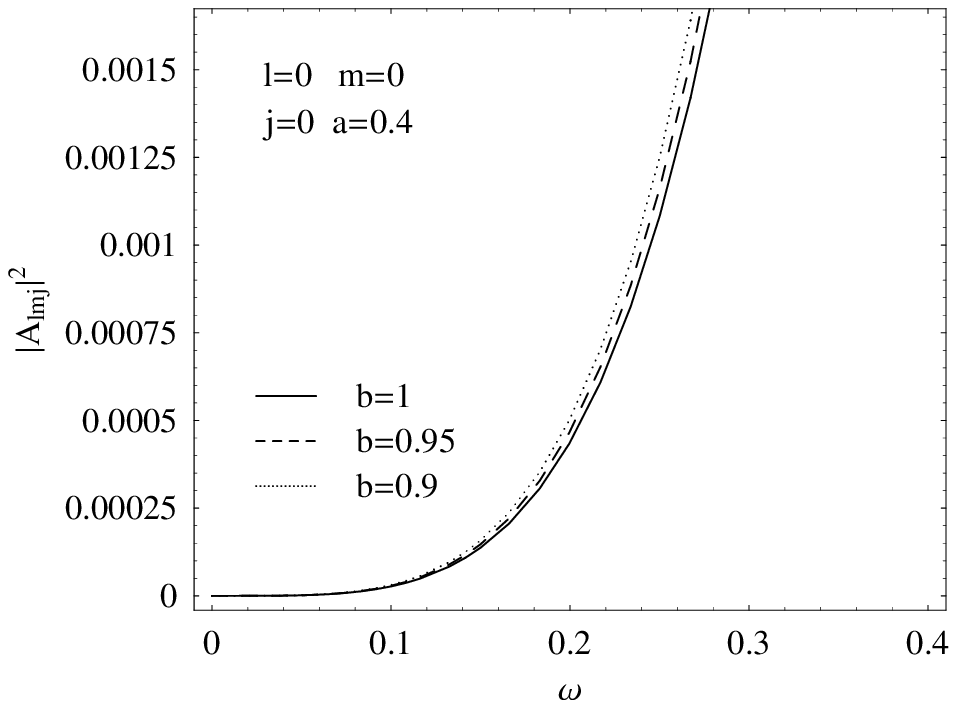}\;\;\;\;\;\;\;\includegraphics[width=8cm]{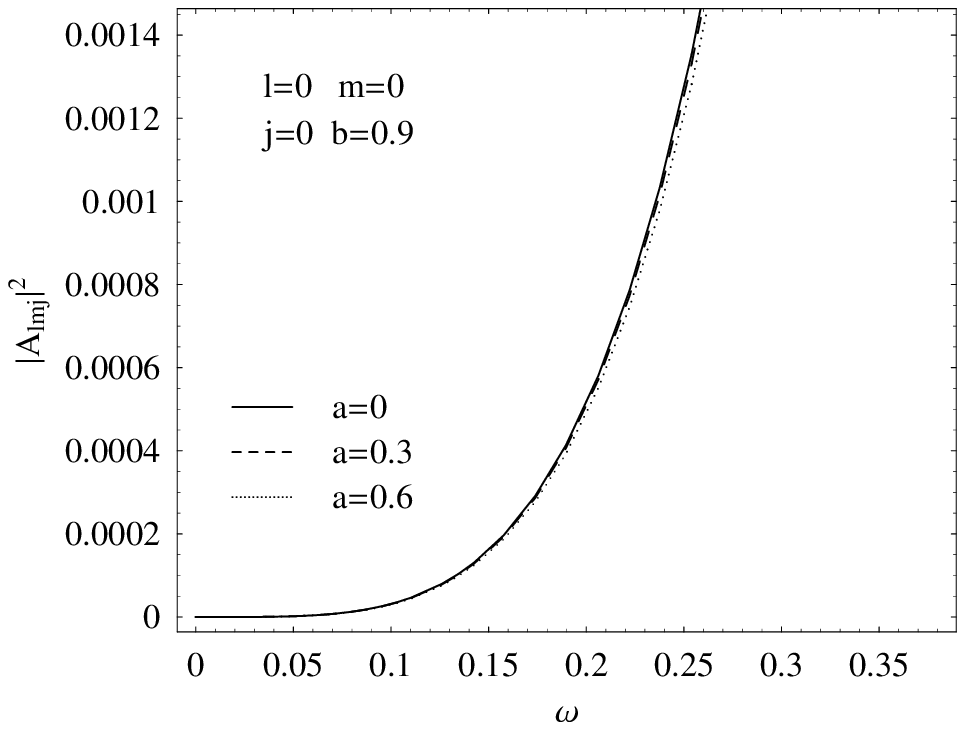}
\caption{Absorption probability $|\mathcal{A}|^2$ of scalar
particles propagating in the rotating black holes on codimension-2
branes, for different $b$ and $a$, respectively, when $l=0$,
$j=0$, $m=0$. Here $r_s=1$.}
\end{center}
\label{fig1}
\end{figure}
\begin{figure}[ht]
\begin{center}
\includegraphics[width=8cm]{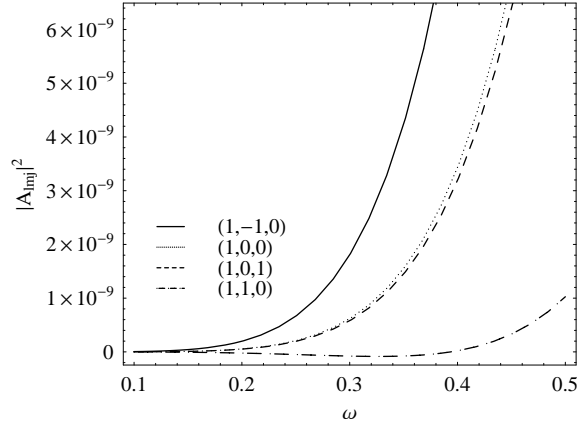} \caption{Absorption
probability $|\mathcal{A}|^2$ of scalar particles propagating in
the rotating black holes on codimension-2 branes for different
combinations of $(l,m,j)$, where we fixed $a=0.4$, $b=0.9$ and
$r_s=1$.}
\end{center}
\label{fig3}
\end{figure}
\begin{figure}[ht]
\begin{center}
\includegraphics[width=8cm]{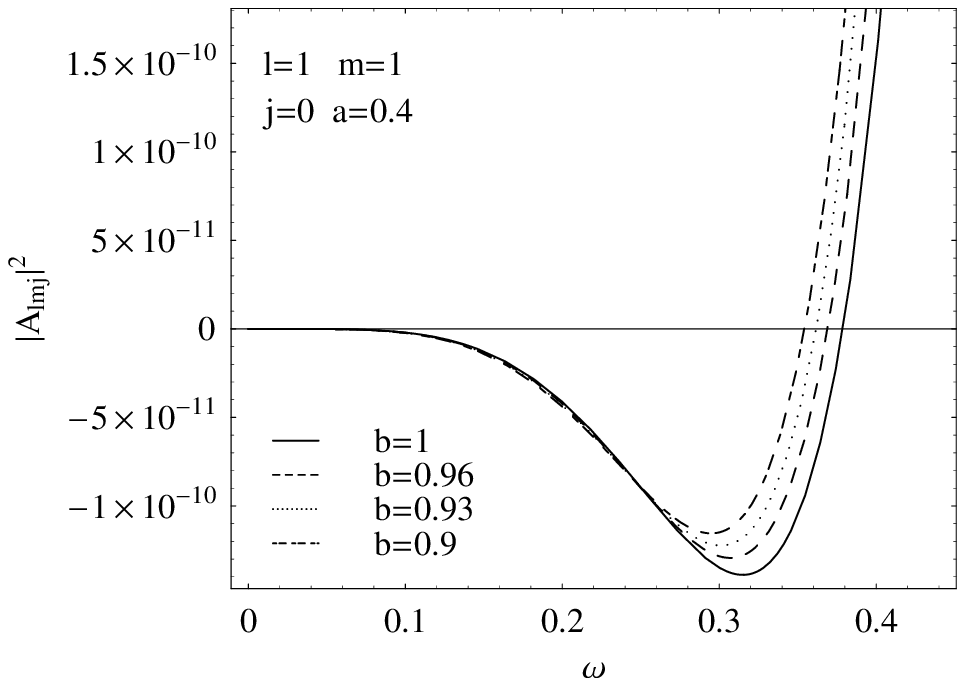}\;\;\;\;\;\;\;\includegraphics[width=8cm]{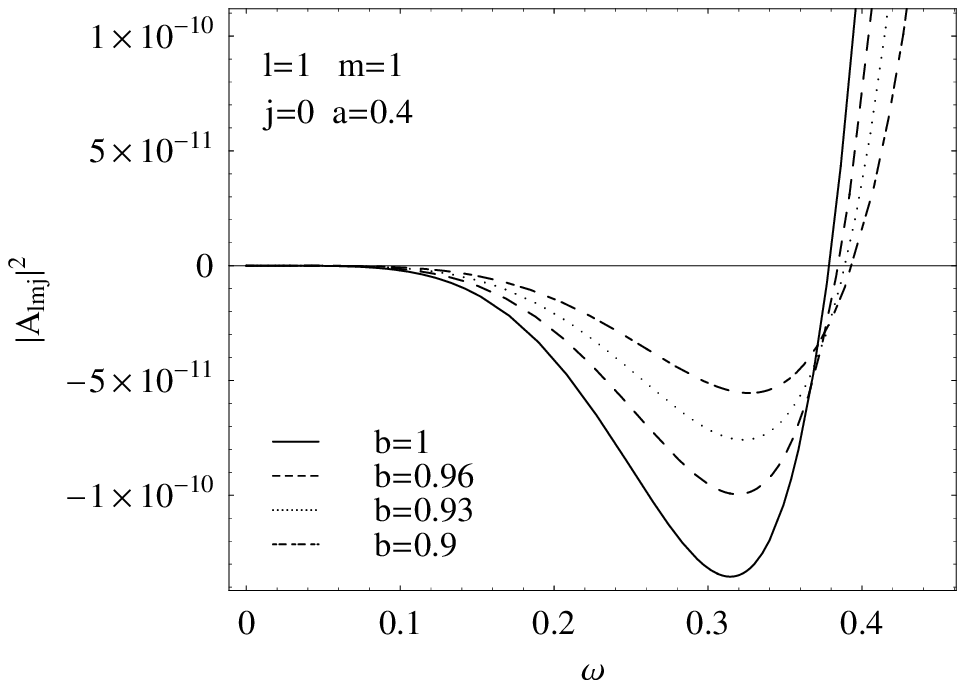}
\caption{Absorption probability $|\mathcal{A}|^2$ of scalar
particles propagating in the rotating black holes (the left for
the metric (\ref{metric1}) and the right for the metric
(\ref{metric2})) on codimension-2 branes, for fixed $a=0.4$,
$l=1$, $m=1$, $j=0$, and different $b$. Here $r_s=1$.}
\end{center}
\label{fig5}
\end{figure}

In Fig.3, we find that for positive $m$, in some ranges of
frequency $\omega$, the absorption probability can be negative,
which presents us the super-radiance. This is the property brought
by the rotation as also disclosed in \cite{9,10}. Here we observed
that the brane tension also influences the super-radiation. For
the black hole spinning orthogonal to the brane, the range of
$\omega$ for the super-radiance to occur increases with the
increase of $b$ (decrease of the brane tension). But for the black
hole spinning on the brane, with the decrease of the brane
tension, the range of $\omega$ for the super-radiance to happen
decreases. The physical reason behind this phenomenon can be
understood as follows.

As in \cite{9,10}, in the low energy limit $BB*\gg i(B*-B)\gg 1$,
we can simplify our (\ref{GFA}) to the form
\begin{eqnarray}
|\mathcal{A}_{lmji}|^2 &=& 2i(\frac{1}{B}-\frac{1}{B*})
\nonumber\\
&=& \frac{4\pi [\omega r^{1/3}_H(r^2_H+a^2)^{1/3}/2 ]^{2l+3}
K_{*,i}\Gamma^2(2\beta+D_*-2)\Gamma^2(1-\beta)(2-D_*-2\beta)}{A(r_H)
\sqrt{E_{lmji}+9/4}\Gamma^2
(\sqrt{E_{lmji}+9/4})\Gamma^2(\beta+D_*-1)\sin^2(\pi(\beta+D_*))}.
\end{eqnarray}
From (33) we learnt that the quantity $2-D_*-2\beta$ is always
positive. The possibility to make $|\mathcal{A}_{lmji}|^2<0$ is
$K_{*,i}<0$, which leads
\begin{eqnarray}
0\leq\omega_1\leq \omega_{c,1}= \frac{am}{r^2_H+a^2},
\end{eqnarray}
for the black hole spinning orthogonal to the brane and
\begin{eqnarray}
0\leq\omega_2\leq
\omega_{c,2}=\frac{am}{b(r^2_H+a^2)},
\end{eqnarray}
for the hole spinning on the brane. Since $r_H\sim r_s b^{-1/3}$,
we have in the low angular momentum limit $\omega_{c,1}\propto
b^{2/3}$ and $\omega_{c,2}\propto b^{-1/3}$, respectively. Thus
with the increase of $b$, $\omega_{c,1}$ increases while
$\omega_{c,2}$ decreases respectively for black holes rotating
along different angles.
\begin{figure}[ht]
\begin{center}
\includegraphics[width=8cm]{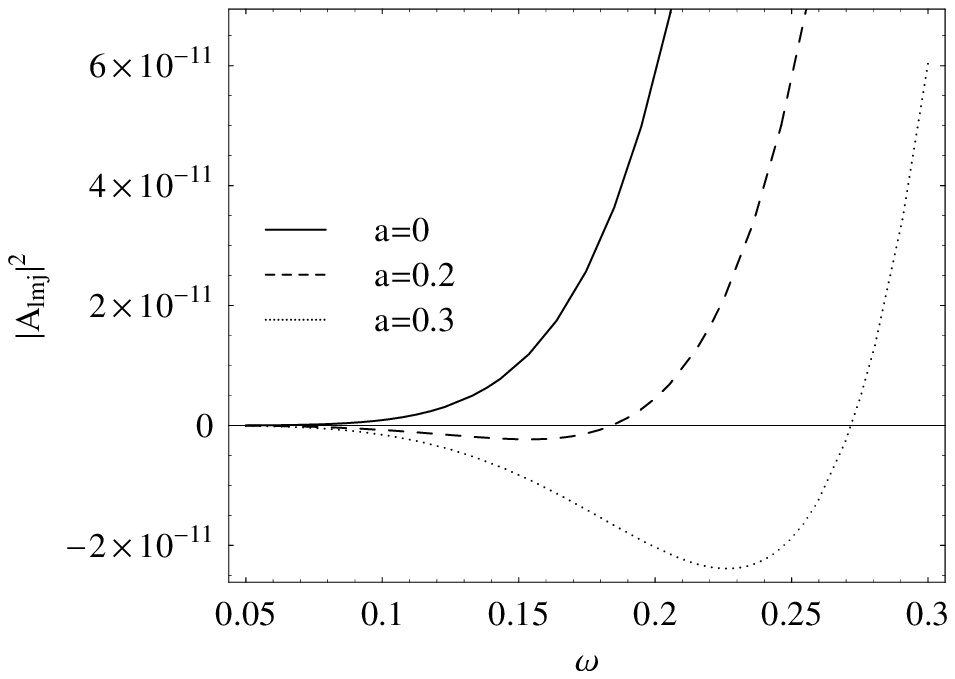}\;\;\;\;\;\;\;\includegraphics[width=8cm]{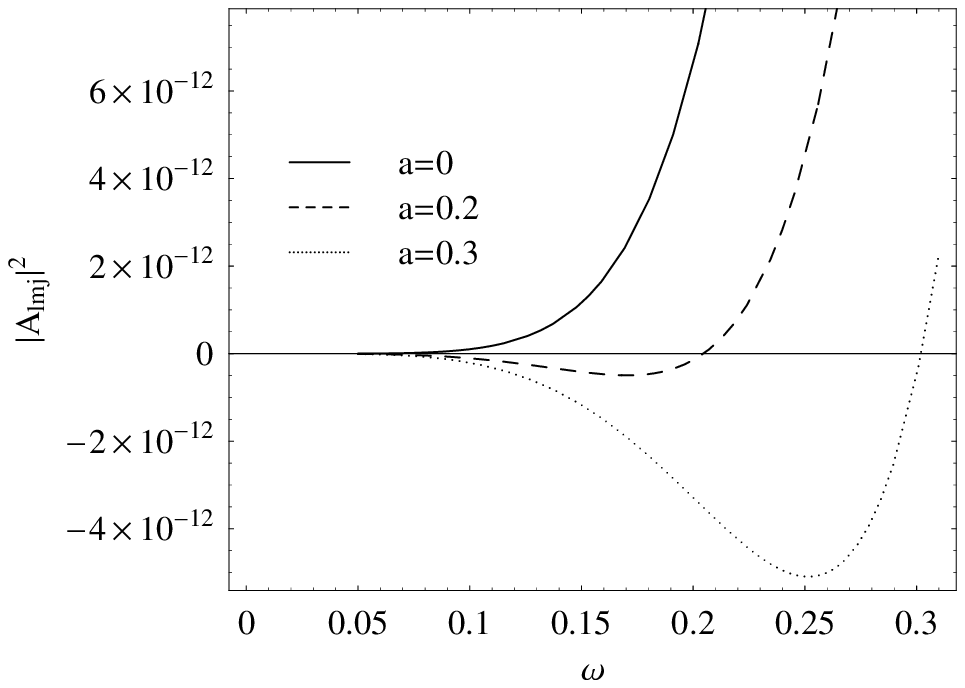}
\caption{The super-radiance for the mode $l=1$,$m=1$,$j=0$ in the
metrics (5) (the left) and (8) (the right) for fixed $b=0.9$ and
different $a$. Here $r_s=1$.}
\end{center}
\end{figure}

The dependence of the super-radiance on the angular momentum of the
black hole is shown in Fig.4. It is clear that the super-radiance
occurs when $a>0$ and becomes stronger when $a$ increases. For the
same angular momentum, it brings more super-radiance in the black
hole background (5) than that in (8). For small $a$ the magnitude of
super-radiance is very small, which has little contribution to the
luminosity of Hawking radiation.

Now let us turn to study the luminosity of Hawking radiation of
black holes in  backgrounds (\ref{metric1}) and (\ref{metric2})
for the mode $l=0$, $m=0$, $j=0$ which plays a dominant role in
the greybody factor. For the first partial wave ($l=0, m=0, j=0$),
we have $K_{*,1}=K_{*,2}$, $P_1=P_2$, $\alpha_1=\alpha_2$ and
$\beta_1=\beta_2$, thus in this case there is no difference in the
absorption probability and Hawking radiation when black holes
rotate along different angles. Performing an analysis similar to
that in \cite{9,10}, we can rewrite the the absorption probability
(50) as
\begin{eqnarray}
|\mathcal{A}_{000}|^2=\frac{4\omega^4
r^2_H(r^2_H+a^2)(3r^2_H+a^2)}{3(9r^2_H+a^2)}.\label{GFA1}
\end{eqnarray}
Combining it with equation (6), the luminosity of Hawking
radiation is given by
\begin{eqnarray}
L&=&\int^{\infty}_0\frac{d\omega}{2\pi}
|\mathcal{A}_{000}|^2\frac{\omega}{e^{\;\omega/T_{H}}-1}\nonumber\\
&=&\frac{(3r^2_H+a^2)^7}{48384\pi r^4_H
(r^2_H+a^2)^5(9r^2_H+a^2)}=\frac{4\pi^5}{63}GT^6_H, \label{LHK}
\end{eqnarray}
\begin{figure}[ht]
\begin{center}
\includegraphics[width=7cm]{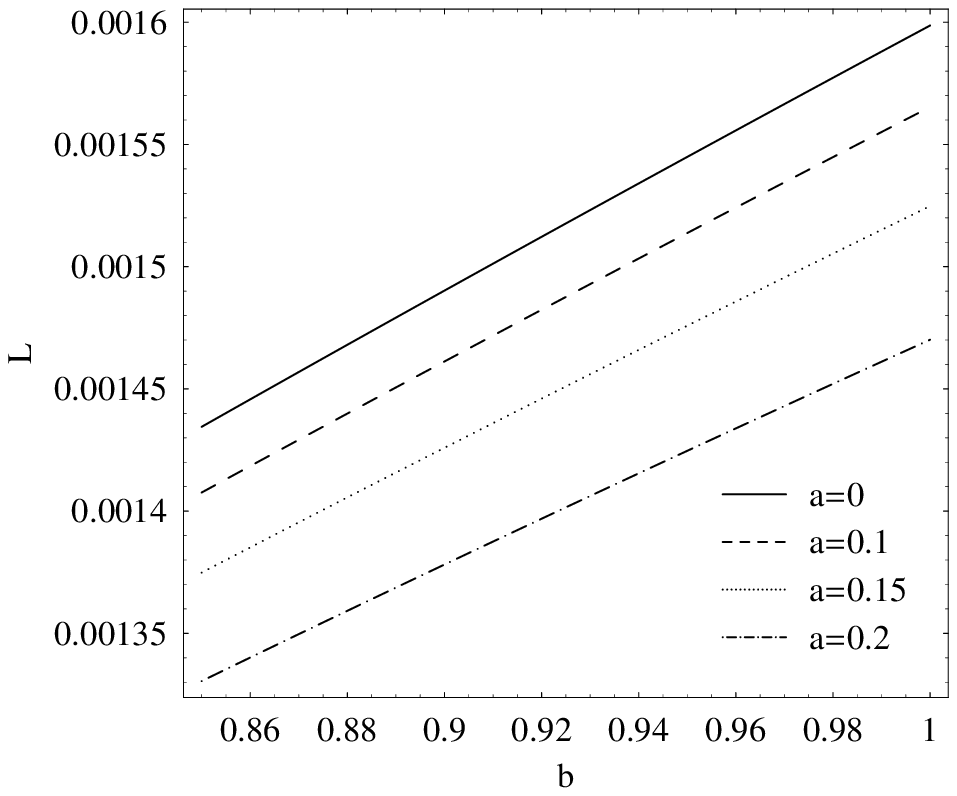}\;\;\;\;\;\;\;\includegraphics[width=8cm]{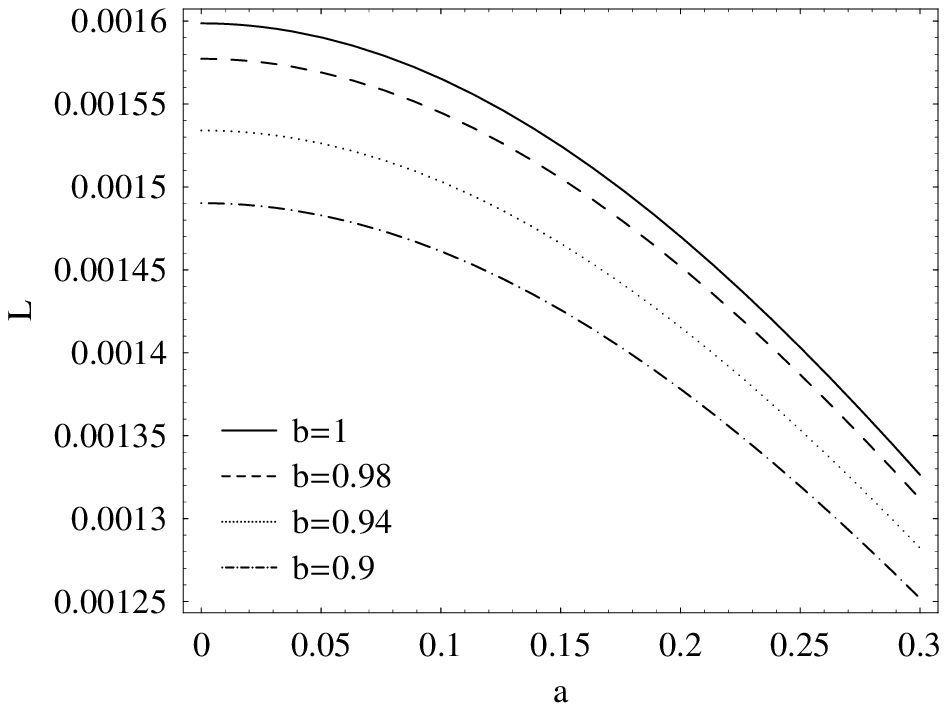}
\caption{The luminosity of Hawking radiation $L$ of scalar
particles propagating in the rotating black holes on codimension-2
branes ($l=1$, $m=1$, $j=0$). The left shows the change of $L$
with $b$ for different $a$ and  the right exhibits the change of
$L$ with $a$ for different $b$. Here $r_s=1$.}
\end{center}
\label{fig6}
\end{figure}
where $G=\frac{4r^2_H(r^2_H+a^2)(3r^2_H+a^2)}{3(9r^2_H+a^2)}$. In
Fig.5, we show the dependence of the luminosity of Hawking radiation
on parameter $b$ for fixed angular momentum parameters. It is clear
that as $b$ increases, (decrease of the brane tension), $L$
increases. From the formula (\ref{LHK}), we obtain that in the low
rotating limit $L\sim r^{-2}_H\sim b^{2/3}r^{-2}_s$. This shows that
the decrease of the brane tension enhances Hawking radiation. This
effect can also be understood from the Hawking temperature. We have
$\frac{dT_H}{db}=\frac{3r^4_H+a^4}{4\pi b^2
r^2_H(r^2_H+a^2)^2(3r^2_H+a^2)}>0$ which indicates that the Hawking
temperature increases with the increase of $b$, thus leading to the
stronger Hawking radiation.

The luminosity of Hawking radiation depending on the angular
momentum of the black hole is also studied. When $a$ increases, the
luminosity of Hawking radiation $L$ decreases. This is because that
the Hawking temperature decreases when the black hole rotates
faster, $\frac{dT_H}{da}=-\frac{a(3r^2_H-a^2)}{2\pi
r_H(3r^4_H+4a^2r^2_H+a^4)}<0$. We have also examined the emission
rate $\frac{dL}{d\omega}=\frac{1}{2\pi}\sum_{jml}\frac{\omega
|\mathcal{A}_{lmji}|^2}{e^{k/T_H}-1}$. We observed that other modes'
influence on the dominant mode $(l=m=j=0)$ is negligible when $a$ is
small. In Figs. 6 and 7,  we show the contribution of other mode on
the dominant mode when $a=1.4$, which still satisfies the low energy
limit $a\omega<1$. Variables $P0$, $P1$, $P2$ in the figure are
given by
\begin{eqnarray}
 P0&=&\frac{1}{2\pi}\frac{\omega
|\mathcal{A}_{000}|^2}{e^{k/T_H}-1},\\
P1&=&\frac{1}{2\pi}\bigg[\frac{\omega
|\mathcal{A}_{110}|^2}{e^{k/T_H}-1}+\frac{\omega
|\mathcal{A}_{1-10}|^2}{e^{k/T_H}-1}+\frac{\omega
|\mathcal{A}_{100}|^2}{e^{k/T_H}-1}\bigg],\\
P2&=&\frac{1}{2\pi}\bigg[\frac{\omega
|\mathcal{A}_{220}|^2}{e^{k/T_H}-1}+\frac{\omega
|\mathcal{A}_{210}|^2}{e^{k/T_H}-1}+\frac{\omega
|\mathcal{A}_{200}|^2}{e^{k/T_H}-1}+\frac{\omega
|\mathcal{A}_{2-10}|^2}{e^{k/T_H}-1}+\frac{\omega
|\mathcal{A}_{2-20}|^2}{e^{k/T_H}-1}\bigg].
\end{eqnarray}
\begin{figure}[ht]
\begin{center}
\includegraphics[width=8cm]{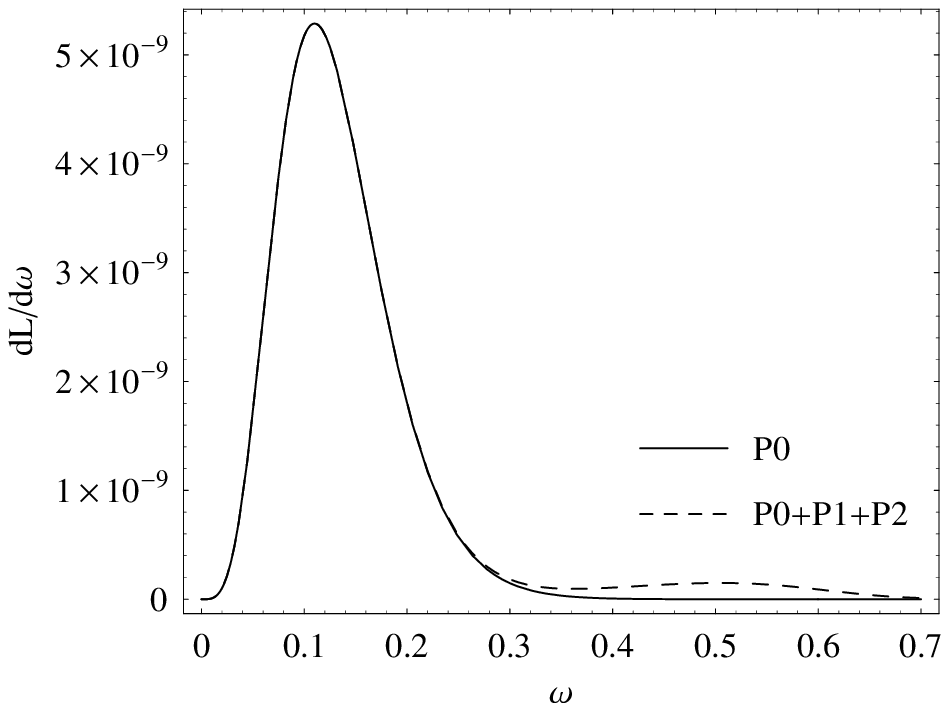}\;\;\;\;\;\;\;\includegraphics[width=8cm]{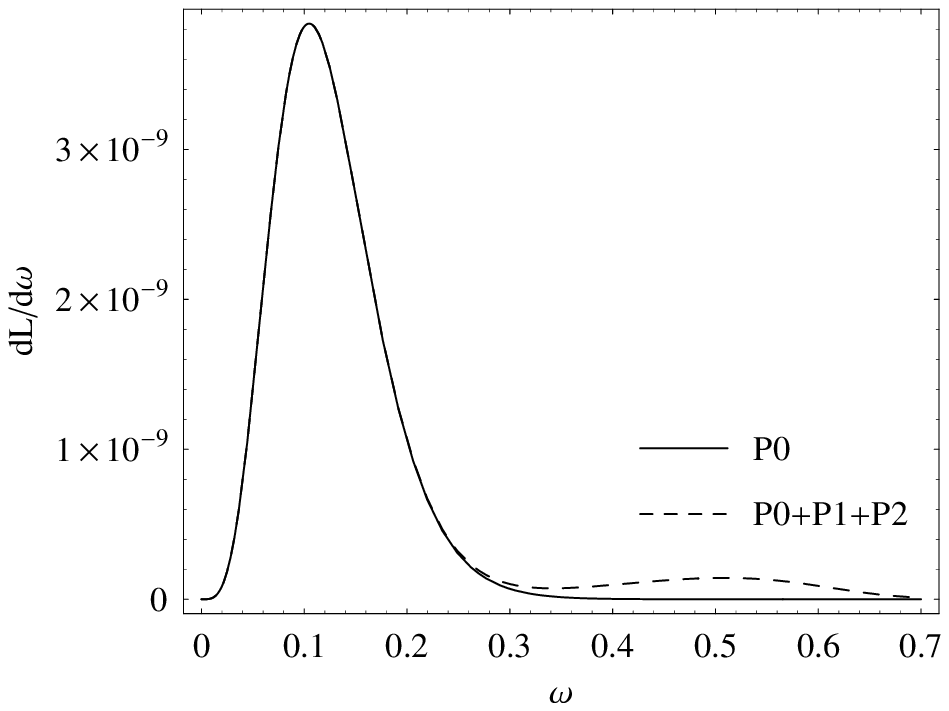}
\caption{$P0$ and $P0+P1+P2$ in metric (5), the left for fixed
$b=0.9$ and the right for fixed $b=0.95$, Here $a=1.4$ and
$r_s=1$. }
\end{center}
\end{figure}
\begin{figure}[ht]
\begin{center}
\includegraphics[width=8cm]{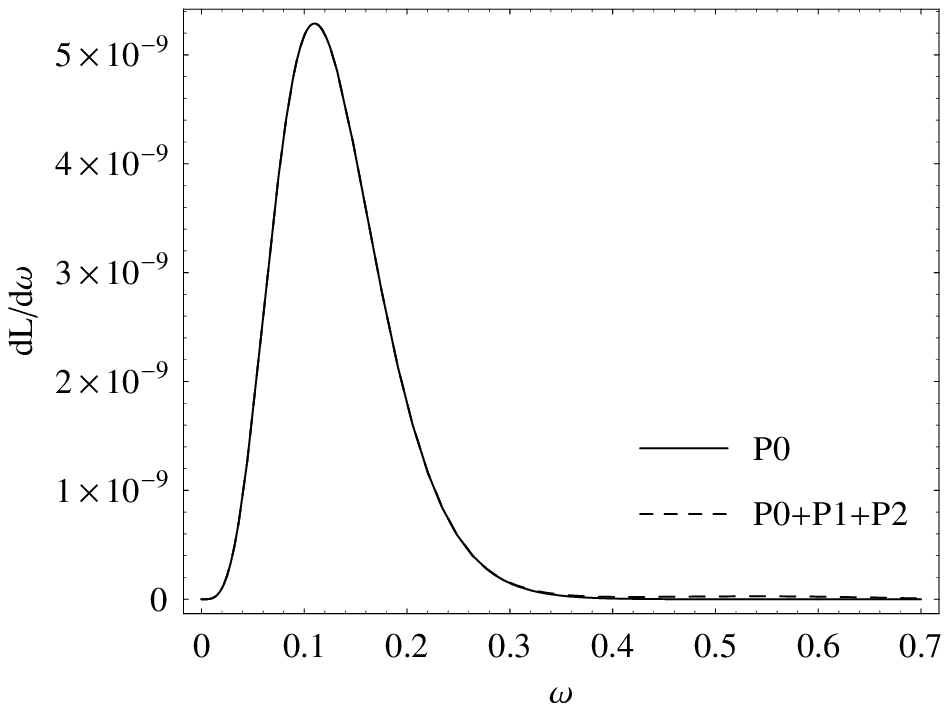}\;\;\;\;\;\;\;\includegraphics[width=8cm]{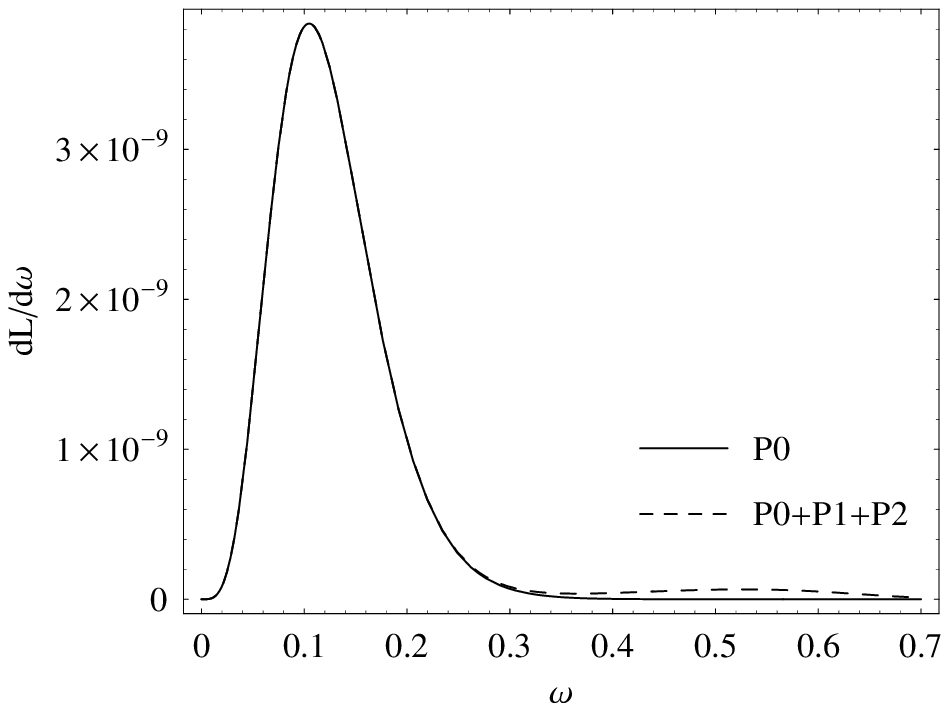}
\caption{$P0$ and $P0+P1+P2$ in metric (8), the left for fixed
$b=0.9$ and the right for fixed $b=0.95$. Here $a=1.4$ and
$r_s=1$.}
\end{center}
\end{figure}
When $b$ increases, the modification becomes bigger. This is
because that bigger $b$ leads stronger super-radiance. Comparing
to the black hole background (5), the modification to the dominant
mode's emission rate is even weaker in the background (8). This is
consistent with the observation that for fixed $a$, there is not
much super-radiance in the background (8) compared to that in (5).
To our observation, in the low energy limit the enhancement of the
emission rate due to the angular momentum and brane tension is not
obvious. It is interesting to generalize our investigation to the
intermediate and high energy and angular momentum situations to
reexamine the emission rate of scalar field.

\section{conclusions and discussions}

We have studied the absorption probability and Hawking radiation
of scalar field in the background of six-dimensional black holes
rotating orthogonal to a tensional brane or spinning on a
tensional brane, respectively. Our results show that with the
nonzero brane tension, properties of evaporations of scalar field
are different from those of the rotating black holes where the
brane tension is completely negligible. This could serve as
signatures of extra dimensions in the future collider searches.

We have observed that the rotation of the black hole brings richer
physics. Nonzero angular momentum triggers the super-radiance. But
in our analytic analysis we see that the enhancement of the emission
rate due to the angular momentum is weak in the low energy limit.
The lowest mode still dominates over other modes in Hawking
radiation and gets small modification in the low energy and angular
momentum limits. It is of interest to generalize our investigation
to the intermediate and high energy and angular momentum situations
to reexamine the emission rate.

We have also compared the results for black holes rotating in
different angles and found that when the black hole rotating on
the tensional brane, the phenomenon of the super-radiance caused
by the angular momentum is not obvious as compared with the black
hole rotating orthogonal to the brane. Less enhancement of the
emission rate has been found for the hole spinning on the
tensional brane. The effects due to the brane tension on the
super-radiance also differ when black holes rotate along different
angles. For the black hole rotating on the tensional brane, the
nonzero brane tension allows bigger range of the frequency to
accommodate the super-radiance.

In this paper we only considered the bulk scalar emission in the
codimension-2 rotating black holes. It would be interesting to
examine the emission of the scalar field on the brane and
investigate the brane to bulk ratio of Hawking radiation. Recently
for rotating black holes with zero tension on the brane, the ratio
between the brane and bulk Hawking radiation has been studied in
\cite{11}. Furthermore it would be of more interesting to study
other fields emission, such as the gravitational field etc. Works in
this direction will be reported in the future.

\newpage
\begin{acknowledgments}

This work was partially supported by NNSF of China, Shanghai
Education Commission and Shanghai Science and Technology
Commission. R. K. Su's work was partially supported by the
National Basic Research Project of China. S. B. Chen's work was
partially supported by the China Postdoctoral Science Foundation
under Grant No. 20070410685, the Scientific Research Fund of Hunan
Provincial Education Department Grant No. 07B043 and the National
Basic Research Program of China under Grant No. 2003CB716300.
\end{acknowledgments}


\begin{thebibliography}{99}
\bibitem{xx} S. B. Giddings and S. Thomas, Phys. Rev. D 65, 056010 (2002);
S. Dimopoulos and G. Landsberg, Phys. Rev. Lett. 87, 161602
(2001); S. Dimopoulos and R. Emparan, Phys. Lett. B 526, 393
(2002); S. Hossenfelder, S. Hofmann, M. Bleicher and H. Stocker,
Phys. Rev. D 66, 101502 (2002)
\bibitem{xx1}  P. Kanti, arXiv:0802.2218.
\bibitem{1} P. Kanti, hep-ph/0310162.
\bibitem{2} C. M. Harris and P. Kanti, JHEP 0310 014 (2003); P. Kanti, Int.
J. Mod. Phys. A 19 4899 (2004).

\bibitem{2a} P. Argyres, S.
Dimopoulos and J. March-Russell, Phys. Lett. B441 96 (1998); T.
Banks and W. Fischler, hep-th/9906038; R. Emparan, G. T. Horowitz
and R. C. Myers, Phys. Rev. Lett. 85 499 (2000).

\bibitem{3}E. Jung and D. K. Park, Nucl. Phys. B 717 272 (2005); N. Sanchez,
Phys. Rev. D 18 1030 (1978); E. Jung and D. K. Park, Class. Quant.
Grav. 21 3717 (2004).

\bibitem{4} A. S. Majumdar, N. Mukherjee, Int. J. Mod.
Phys. D 14 1095 (2005) and reference therein; G. Kofinas, E.
Papantonopoulos and V. Zamarias, Phys. Rev. D 66, 104028 (2002);
G. Kofinas, E. Papantonopoulos and V. Zamarias, Astrophys. Space
Sci. 283, 685 (2003); A. N. Aliev, A. E. Gumrukcuoglu, Phys. Rev.
D 71, 104027 (2005); S. Kar, S. Majumdar, hep-th/0510043; S. Kar,
S. Majumdar, hep-th/0606026; S. Kar, hep-th/0607029.

\bibitem{5} V. P. Frolov, D. Stojkovic, Phys. Rev. Lett. 89, 151302
(2002); Valeri P. Frolov, Dejan Stojkovic, Phys. Rev. D 66, 084002
(2002); D. Ida, K. Oda and S. C. Park, Phys. Rev. D 67 064025
(2003);  D. Stojkovic, Phys. Rev. Lett. 94, 011603 (2005); E.
Jung, S. H. Kim and D. K. Park, Phys. Lett. B 615 273 (2005); E.
Jung, S. H. Kim and D. K. Park, Phys. Lett. B 619 347 (2005);  G.
Duffy, C. Harris, P. Kanti and E. Winstanley, JHEP 0509 049
(2005); E. Jung and D. K. Park, Nucl. Phys. B 731 171 (2005); M.
Casals, P. Kanti and E. Winstanley, JHEP 0602 051 (2006); A. S.
Cornell, W. Naylor and M. Sasaki, JHEP 0602, 012 (2006).


\bibitem{6} D.K. Park, hep-th/0512021; Eylee Jung and D. K. Park,
hep-th/0612043; V. Cardoso, M. Cavaglia, L. Gualtieri, Phys. Rev.
Lett. 96, 071301 (2006); V. Cardoso, M. Cavaglia, L. Gualtieri,
JHEP 0602, 021 (2006).

\bibitem{7}D. Dai, N. Kaloper, G. Starkman and D. Stojkovic, Phys.Rev. D75 (2007) 024043, hep-th/0611184.

\bibitem{8} L.H. Liu, B. Wang, G.H. Yang, hep-th/0701166, Phys.Rev.D76, 064014 (2007).

\bibitem{9} S. Creek, O. Efthimiou, P. Kanti and K. Tamvakis, Phys. Rev. D 75 084043 (2007).

\bibitem{10} S. B. Chen, B. Wang, R. K. Su, arXiv:0710.3240.

\bibitem{11}  S. Creek, O. Efthimiou, P. Kanti, K. Tamvakis,
arXiv:0709.0241; arXiv:0707.1768.

\bibitem{12} J. Y. Shen, B. Wang and R. K. Su, Phys. Rev. D 74, 044036 (2006).

\bibitem{13} E. Abdalla, B. Cuadros-Melgar, A. B. Pavan and C. Molina,
Nucl.Phys. B 752, 40 (2006).

\bibitem{14} S. B. Chen, B. Wang, R. K. Su, Phys. Lett. B 647, 282 (2007),
arXiv:hep-th/0701209.

\bibitem{15} P. Kanti, R. A. Konoplya, A. Zhidenko, gr-qc/0607048 ; P. Kanti,
R. A. Konoplya, Phys. Rev. D 73, 044002 (2006); D. K. Park, Phys.
Lett. B 633, 613 (2006).

\bibitem{16} U. A. al-Binni, G. Siopsis, arXiv:0708.3363;  H. T. Cho, A. S. Cornell, J. Doukas, W. Naylor,
arXiv:0710.5267; arXiv:0709.1661.

\bibitem{17} N. Kaloper and D. Kiley, JHEP 0603 077 (2006).

\bibitem{18} D. Kiley, arxiv: 0708.1016.

\bibitem{19}  T. Kobayashi, M. Nozawa, Y. Takamizu,
arXiv:0711.1395.

\end{thebibliography}
\end{document}